\documentclass[structabstract]{aa}

\def\gsim{ \lower .75ex \hbox{$\sim$} \llap{\raise .27ex \hbox{$>$}} }
\def\lsim{ \lower .75ex\hbox{$\sim$} \llap{\raise .27ex \hbox{$<$}} }
\usepackage{graphicx}
\usepackage{txfonts}
\usepackage{lscape}
\begin{document}

\title{FERMI constraints on the high energy, $\sim 1$GeV, emission of long GRBs}

\author{Dafne Guetta\inst{1},  Elena Pian\inst{2,3,4}  and Eli Waxman\inst{5}}

\institute{Osservatorio astronomico di Roma, v. Frascati 33,
00040 Monte Porzio Catone, Italy \and
Osservatorio Astronomico di Trieste, Via G.B. Tiepolo, 11 - 34143 Trieste, Italy \and
Scuola Normale Superiore, Piazza dei Cavalieri 7, I-56126 Pisa, Italy \and
European Southern Observatory, Karl-Schwarzschild-Strasse 2
D-85748 Garching bei M\"unchen, Germany \and
Dept. of Particle Phys. \& Astrophys., Weizmann Institute of Science, Rehovot 76100, Israel }

%\date{Received September 15, 1996; accepted March 16, 1997}

\abstract{}
{We investigate the constraints imposed on the luminosity function (LF) of long duration Gamma Ray Bursts (LGRBs) by the flux distribution of bursts detected by the GBM at $\sim1$~MeV, and the implications of the non detection of the vast majority, $\sim$95\%, of the LGRBs at higher energy, $\sim1$~GeV, by the LAT detector. }
{We find a LF that is consistent with those determined by BATSE and {\it Swift}. The non detections by LAT set upper limits on the ratio $R$ of the prompt fluence at $\sim1$~GeV to that at $\sim1$~MeV. The upper limits are more stringent for brighter bursts, with $R<\{0.1,0.3,1\}$ for $\{5,30,60\}\%$ of the bursts. This implies that for most bursts the prompt $\sim1$~GeV emission may be comparable to the $\sim1$~MeV emission, but can not dominate it. The value of $R$ is not universal, with a spread of (at least) an order of magnitude around $R\sim10^{-1}$.
For several bright bursts with reliable determination of the photon spectral index at $\sim1$~MeV, the LAT non detection implies an upper limit to the $\sim100$~MeV flux which is $<0.1$ of the flux obtained by extrapolating the $\sim1$~MeV flux to high energy.}
{ For the widely accepted models, in which the $\sim1$~MeV power-law photon spectrum reflects the power-law energy distribution of fast cooling electrons, this suggests that either the electron energy distribution does not follow a power-law over a wide energy range, or that the high energy photons are absorbed. Requiring an order unity pair production optical depth at 100~MeV  sets an upper limit for the Lorentz factor, $\Gamma\lsim10^{2.5}$.}
{}

\keywords{gamma rays: bursts }

\maketitle

\section{Introduction}

Gamma ray bursts (GRBs) are the most powerful explosions in the universe, with apparent (isotropic equivalent) energy output sometimes exceeding $10^{54}$ ergs. While it is widely accepted that GRBs are produced by the dissipation of energy in highly relativistic winds driven by compact objects (see, e.g., M\'esz\'aros 2006, Piran 2004 and Waxman 2003 for reviews) the physics of wind generation and radiation production is not yet understood. It is not known, for example, whether the wind luminosity is carried, as commonly assumed, by kinetic energy or by Poynting flux (e.g. Drenkhahn \& Spruit  2002, Lyutikov et al.  2003), whether the radiating particles are accelerated by the dissipation of magnetic flux or by internal shocks dissipating kinetic energy, and whether the emission is dominated by synchrotron or inverse-Compton radiation of accelerated electrons, as commonly assumed, or by hadronic energy loss of accelerated protons (see  Dermer \& Fryer  2008 and references therein).  

GRBs were detected by instruments sensitive mainly in the 100 to 1000 keV range like BATSE  (Paciesas et al. 1999)  and the GRBM on BeppoSAX (Guidorzi et al. 2004). Measurements at high energy were limited by the lower sensitivity and/or the smaller field of view of higher energy instruments (e.g. CGRO/EGRET- Dingus 1995, Hurley et al. 1994, Gonzalez et al. 1994; AGILE/GRID-Marisaldi et al. 2009, Giuliani et al. 2008,  Giuliani et al. 2010). The improved high energy, $\gsim1$~GeV, sensitivity and field of view of the instruments on board the Fermi satellite (Atwood et al. 2009, Band et al. 2009) are expected to improve the quantity and quality of high energy GRB data, and may therefore provide qualitatively new constraints on models.

The main goal of this paper is to determine the implications of the non-detection of the vast majority of long duration ($T>2$~s) GRBs by  Fermi's LAT detector. We first show in \S~\ref{sec:LF} that the LF of LGRBs detected by Fermi's GBM at $\sim1$~MeV is consistent with those inferred from BATSE and {\it Swift} observations, and that these instruments sample the LF in a similar manner. We then derive in \S~\ref{sec:non_detect} the constraints on the $>100$~MeV emission implied by LAT non-detections. In \S~\ref{sec:EGRET} we compare our results with those inferred from the analysis of EGRET data (Gonzalez Sanchez 2005). Our results are summarized and their implications are discussed in \S~\ref{sec:discussion}.

\section{The luminosity function  of the GBM sample}
\label{sec:LF}

\subsection{The GBM, BATSE and {\it Swift}-BAT samples}
\label{sec:samples}

Our  methodology follows that of Guetta et al. (2005) and of Guetta \& Piran (2007). We consider all long ($T_{90} > 2$ seconds) bursts detected by the GBM until February 2010 (see Table 2) and compare the distribution of their peak fluxes with that of LGRBs detected by BATSE and {\it Swift}. Since the energy dependent sensitivity of the different instruments is different, we convert the GBM fluxes to equivalent fluxes in the 50-300~keV band, which is the band used by BATSE for GRB triggers. This conversion is carried out assuming that the energy spectrum around the light curve peak is well described by the spectral model that has been adopted to fit the time average spectrum. The spectral model parameters have been collected from the literature (mainly GCN circulars) and are reported in Table 2. 

We have excluded from our analysis GRB081126, which  has 2 pulses, because the measurements reported in GCN 8589 are inconsistent: the peak flux and fluence of the event are reported to be smaller than those reported for the individual pulses.  We have kept GRB090423 in our analysis, despite the debate regarding its classification as a long burst (see Salvaterra et al. 2009) since its duration does satisfy $T_{90}>2$~s.

For bursts reported in Table 2 with spectra fitted by a "Band law", $dN_\gamma/dE\propto E^{\alpha}$ for $E<E_p$ and $dN_\gamma/dE\propto E^{\beta}$ for $E>E_p$, we find $<\alpha>\simeq -0.8$, $<\beta>\simeq -2.3$, and $<E_p>\simeq 200$~keV. Using these average parameters, we estimate, based on Band (2003), a GBM sensitivity in the 50-300~keV band of $P^{(50-300)\rm keV}_{\rm lim, GBM}\sim$ 0.6 ph cm$^{-2}$ s$^{-1}$. For our LF analysis we use only the 144 long ($T_{90} > 2$~s) GBM bursts with peak flux higher than that threshold.

In the BATSE sample we have included all LGRBs detected while the BATSE onboard trigger was set to a significance of 5.5~$\sigma$ over background in at least two detectors in the energy range of 50-300~keV. Among those we selected the bursts for which the peak flux in 1024~ms time bins is higher than the BATSE threshold for long bursts reported by Band (2003), $P^{(50-300)\rm keV}_{\rm lim, BATSE}\sim$ 0.25 ph cm$^{-2}$ s$^{-1}$. This yields a sample of 1425 bursts.

For {\it {\it Swift}} we consider LGRBs detected until September 2009 in the energy range 15-150~keV. We convert the BAT 15-150~keV peak fluxes to fluxes in the 50-300~keV band using the BAT peak fluxes and spectral parameters provided by the {\it Swift} team\footnote{See {\it {\it Swift}} information page\\
 http://{\it Swift}.gsfc.nasa.gov/docs/{\it Swift}/archive/grb\_table.html}. We consider only the bursts with peak fluxes above the threshold of $P^{(50-300)\rm keV}_{\rm lim, {\it Swift}}\sim 0.3$ ph cm$^{-2}$ s$^{-1}$ (Gorosabel et al. 2004, Band 2003), yielding a sample of 259 GRBs.

The flux distributions of bursts included in the samples described above are shown in figure 1.

\begin{figure}
\label{fig:logNlogS}
{\par\centering \resizebox*{0.95\columnwidth}{!}
{\includegraphics
{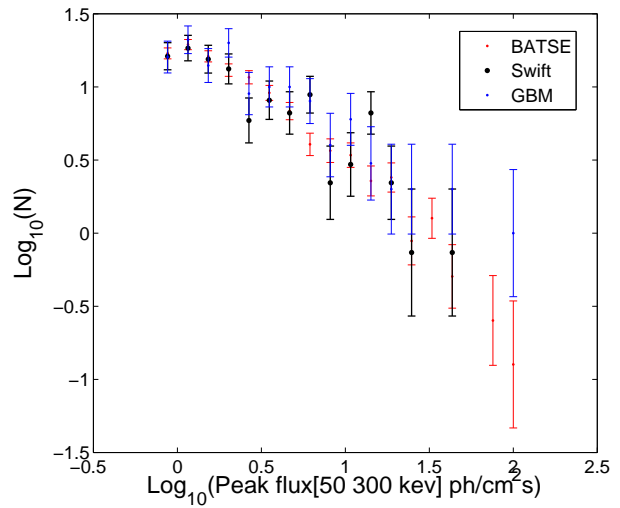}} \par} \caption{The flux distributions of bursts included in the BATSE, {\it Swift} and GBM samples (see text). Shown are the number of bursts detected within 20 equally spaced intervals of log peak flux (in the 50-300~keV band). Error bars derived assuming Poisson statistics.}
\end{figure}

\subsection{Luminosity function and comoving rate}
\label{sec:LF_rate}

The method used to derive the luminosity function is essentially the same as that used by Schmidt (1999), Guetta et al. (2005) and Guetta and Piran (2007). We consider a broken power law luminosity function,
\begin{equation}
\label{Lfun}
\Phi(L)\equiv\frac{dN}{d\log L}\propto
\left\{ \begin{array}{ll}
(L/L^*)^{-a_1} &  L^*/\Delta_1 < L < L^* \\
(L/L^*)^{-a_2} & L^* < L < \Delta_2 L^*
\end{array}
\right. \;,
\end{equation}
with $\Delta_1=100$, $\Delta_2=100$ and $\{L^*,a_1,a_2\}$ independent of redshift, and a (comoving) rate evolving with redshift following the  Porciani \& Madau (2001) star formation rate,
\begin{equation}
\label{SFR} 
R_{GRB}(z)\propto \frac{\exp(3.4z)}{\exp(3.4z)+22}.
\end{equation}
The cosmological "k correction" is determined adopting an effective spectral index in the observed range of 50-300~keV band of -1.6 ($dN_\gamma/dE\propto E^{-1.6}$) (see, e.g., Guetta and Piran 2007).

In order to determine $\{L^*,a_1,a_2\}$ we use Monte Carlo simulations to generate an ensemble of distributions of burst peak fluxes, that would have been detected by the various instruments for a given set of $\{L^*,a_1,a_2\}$ values. We find the best-fitting LF parameters $\{L^*,a_1,a_2\}$ and their uncertainty by $\chi^2$ minimization. The results are reported in Table 1. The LF parameters derived from the different samples are consistent with each other. Note, that the luminosity considered here is the "isotropic" equivalent luminosity.

\begin{table}[h]
\label{table:LF}
\caption{Best fit LF parameters$^1$}
\begin{tabular}{|c|c|c|c|c|c|}
 \hline
sample
& Rate(z=0)\tablefootmark{1}   & $L^*$ [50-300]~keV & a$_1$ & a$_2$ &$\chi^2$/D.O.F.\footnotemark[3]\\
   & ${\rm Gpc^{-3} yr^{-1}}$ & $10^{51}$~erg/s &  & &   \\
  \hline
GBM & $0.5^{+0.3}_{-0.2}$ & $5.5^{+1.5}_{-2}$ & $0.3^{+0.1}_{-0.5}$
&$2.3^{+0.6}_{-0.3}$ & 1.1  \\
BATSE & $1.0^{+0.2}_{-0.4}$ & $4_{-1.5}^{+2}$ &
$0.1_{-0.1}^{+0.3}$ &$2.6_{-0.5}^{+0.9}$ & 1.1 \\
{\it {\it Swift}} & $0.6^{+0.3}_{-0.1}$ & $3.3_{-0.5}^{+2.5}$ &
$0.1_{-0.1}^{+0.3}$ &$2.7_{-0.4}^{+1}$ & 0.95 \\
  \hline
\end{tabular}
\tablefoot{
\tablefoottext{1}{Best fit values and their 1~$\sigma$ uncertainty range.}\\
\tablefoottext{2}{Integrated over $L$.}\\
\tablefoottext{3}{$\chi^2$ obtained for best fit values.}
}
\end{table}

\section{Constraints on the $>100$~MeV emission implied by LAT non-detections}
\label{sec:non_detect}

Out of the 205 LGRBs detected by the GBM, only a few, $\sim$~12 bursts, have been detected above 30 MeV by the LAT telescope. Figures~2 and~3 present the upper limits implied by the LAT non-detection on the ratio of the prompt emission 
fluences at 100~MeV and 1~MeV. 
We have included in the burst sample shown in the figure the 121 LGRBs for which an accurate determination of the 1~MeV fluence is possible  based on the fluence measured by the 
Fermi GBM and on reliable spectral fit parameters (see table~2)\footnote{Some GRBs have reported fluence and no reported peak flux. While these GRBs were excluded from the peak flux analysis described in \S~\ref{sec:LF}, they have been included in the fluence ratio analysis.}$^,$\footnote{For 4 GRBs with multi-peak structure, separate sets of spectral fit parameters are reported in the literature for each pulse (GRBs 081009, 090509, 090516A, 090610B). In these cases, we have determined the 1 MeV fluence using the spectrum of the second pulse (see  Table 2), under the assumption that the bulk of the MeV-GeV output is emitted simultaneously with the second pulse. This is motivated by the 2 long GRBs with detailed published GBM and LAT light curves, GRB080916C (Abdo et al. 2009)  and GRB090902B (Bissaldi et al. 2009).}$^,$\footnote{
For GRBs with spectra fitted in the GBM band with single power-laws, the spectrum cannot be extrapolated straightforwardly to 1~MeV if the energy range used for the power-law fit is much softer. Thus, we have retained the GRBs fitted with single power-laws over an energy reaching or exceeding 800 keV, and excluded from the analsyis GRBs 080818A, 080928, 081206C, 081225.}, and for which the reported LAT bore sight angle is less than $80^\circ$ (as the LAT effective area is very small for bursts observed at larger angles). The upper limit on the $>100$~MeV fluence implied by the LAT non-detection is derived as follows.

\begin{figure}[h]
\label{fig:R2}
{\par\centering \resizebox*{0.95\columnwidth}{!}
{\includegraphics
{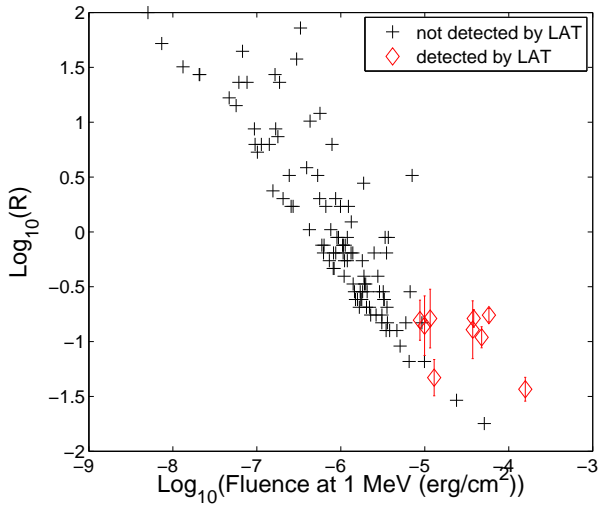}} \par} \caption{The upper limits implied by the LAT non-detections on the ratio $R$ of the 100~MeV to 1~MeV fluences, 
$\nu F_\nu=\nu\int dt f_\nu(t)$ for $h\nu=1,100$~MeV, 
during the prompt emission, assuming a differential photon spectrum $dN_\gamma/dE\propto E^{-2}$ above $100$~MeV. Also shown (red diamonds) are the measured ratios for GRBs detected by the LAT (GRB080825C, GRB080916C, GRB090217,  GRB090323, GRB090328, GRB090626, GRB090902B, GRB090926A,  GRB091003,  reported in Ghisellini et al. 2010 and Abdo et al. 2010). The 100~MeV specific fluence was derived for these bursts from the reported $>100$~MeV fluence assuming $dN_\gamma/dE\propto E^{-2}$ above $100$~MeV.}
\end{figure}
A GRB is tagged as detected by the LAT if the number of photons detected, $N_s$, exceeds 10 and exceeds a $5\sigma$ fluctuation of the background (Band et al. 2009, Atwood et al. 2009). For the current analysis, it is sufficient to consider the $N_s>10$ criterion, since the number of background events detected during the characteristic time of the prompt $\sim1$~MeV gamma-ray emission, $T_{90}\lsim 100$~s, is low ($\sim1$, i.e. $N_s>10$ is above a 5$\sigma$ background fluctuation). Following Band et al. (2009), the expected number of counts from a burst with a time integrated differential photon flux (i.e. differential photon fluence) $Q(E)$ is
\begin{equation}
N_S = \int_{E_1}^{E_2}dE A_{eff}(E,\theta)Q(E),
\label{ns}
\end{equation}
where $A_{eff}(E,\theta)$ is the effective area (taken from Atwood et al. 2009) that depends on the direction from which the burst is observed, $\theta$, $E_1=100$ MeV and $E_2=10$ GeV.

\begin{figure}[h]
\label{fig:R3}
{\par\centering \resizebox*{0.95\columnwidth}{!}
{\includegraphics
{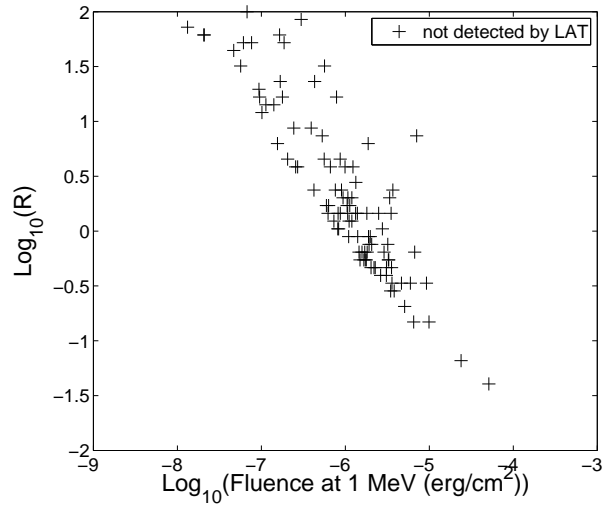}} \par} \caption{Same as figure~2, assuming a differential photon spectrum $dN_\gamma/dE\propto E^{-3}$ above $100$~MeV.}
\end{figure}

The upper limits on the $100$~MeV fluences shown in figures~2 and~3 are obtained by requiring $Q(E)<Q_0(E)$ where $Q_0(E)$ is the fluence for which $N_s=10$. Since the number of events detected for bursts with fluence that produces, on average, $N_s=10$ events is Poisson distributed with an average equal to $10$, the upper limits on the fluxes shown in the figure are $\sim50\%$ confidence level upper limits. The upper limits on the fluence for confidence levels of 70\% and 99.7\% are 1.2 and 2 times higher respectively.

\begin{figure}[h]
\label{fig:histR}
{\par\centering \resizebox*{0.95\columnwidth}{!}
{\includegraphics
{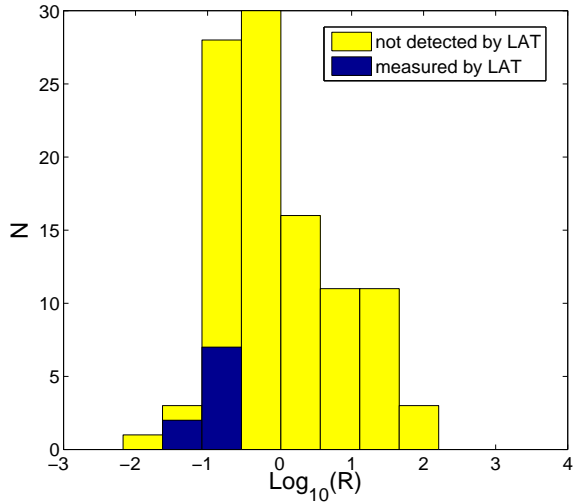}} \par} \caption{A histogram representation of the upper limits on, and measured values of, $R$, presented in fig.~2.}
\end{figure}

The effective area of the LAT is roughly proportional to $E$ in the energy range of 100~MeV to 1~GeV (and roughly energy independent between 1~GeV and 10~GeV). This implies that the LAT flux sensitivity is roughly energy independent between 100~MeV and 1~GeV, and that the upper limit on $N_s$ implies an upper limit on the 0.1--1~GeV fluence, which is independent of the spectrum $Q(E)$. Indeed, the upper limits on $R$, the 100~MeV to 1~MeV fluence ratio, obtained assuming $Q(E)\propto E^{-3}$ are higher by a factor of $\sim2$ than those obtained for a $Q(E)\propto E^{-2}$, implying that the upper limits on the 0.1--1~GeV fluence obtained for both spectral shapes are similar. Denoting by $R^{(2)}$ the upper limit obtained on $R$ assuming $Q(E)\propto E^{-2}$, the upper limit on the 0.1--1~GeV fluence is $\approx\ln(10)R^{(2)}\nu F_\nu|_{1\rm MeV}$, the upper limit on the 1--10~GeV fluence is few times higher, and the upper limit on the ratio of the 0.1--1~GeV fluence and the 0.1--1~MeV fluence is $\approx R^{(2)}$. We find $R^{(2)}<\{0.1,0.3,1\}$ for $\{5,30,60\}\%$ of the bursts respectively. Figure~4 presents a histogram of the distribution of upper limits on $R^{(2)}$.

\begin{figure}[h]
{\par\centering \resizebox*{0.95\columnwidth}{!}
{\includegraphics
{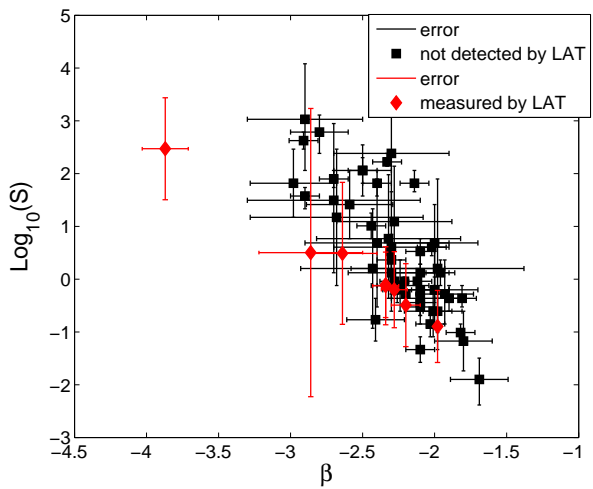}} \par} \caption{Upper limits implied by LAT non-detections on the ratio $S$ between the 
100~MeV fluence and the 100~MeV fluence obtained by extrapolation to 100~MeV of the $\sim1$~MeV spectrum of the GBM detected photons. $\beta$ is the spectral index describing the high energy part of the GBM spectrum, $dN_\gamma/dE\propto E^\beta$. Also shown are the measured values of $S$ for LAT detected GRBs (red).}
\end{figure}

\begin{figure}[h]
{\par\centering \resizebox*{0.95\columnwidth}{!}
{\includegraphics
{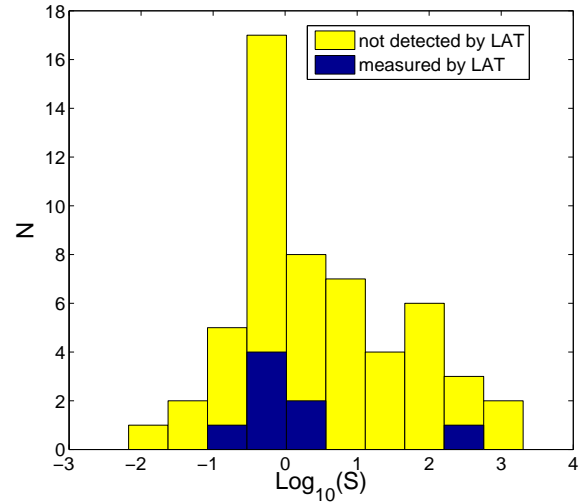}} \par} \caption{A histogram representation of the results of fig.~5.}
\end{figure}

In figure~5 we present the upper limits implied by the LAT non-detection on the ratio $S$ between the 
100~MeV fluence and the 100~MeV fluence obtained by extrapolation to 100~MeV of the $\sim1$~MeV spectrum of the GBM detected photons. For this comparison, we have used only the $\sim60$ bursts with spectra which are well fitted by a "Band function" and for which the bore sight angle is $<80^\circ$. We have excluded from the sample shown in the figure bursts for which the GBM spectrum is not fitted by a Band law, i.e. fitted by either 
a power-law with exponential suppression (which predicts negligible flux at 100~MeV) or by a single power-law (for which a spectral break above 1~MeV is not unexpected). 
A histogram representation of the results of fig.~5 is shown in fig.~6. The figures show that for at least a few bursts the non detection by LAT implies a significant suppression of the 100~MeV flux compared to that expected from an extrapolation to high energy of the $\sim1$~MeV power-law spectrum.

\section{Comparison with EGRET's results}
\label{sec:EGRET}

We compare in this section our results with those obtained by Gonz\'alez S\'anchez (2005) and Ando et al. (2008) from the non detection by EGRET of bursts detected by BATSE. EGRET detected photons in the 30~MeV-30~GeV energy range, with effective area roughly proportional to $E$ between 30~MeV and 200~MeV, and approximately energy independent ($\approx 10^3$~cm$^2$) at higher energy (see Thompson et al. 1993). This area is similar to the LAT effective area at $\sim100$~MeV and smaller by roughly an order of magnitude than the LAT effective area at $\sim1$~GeV. From table~2.1 and fig.~2.3 of Gonz\'alez S\'anchez (2005), EGRET's non-detections imply upper limits on the ratio of the EGRET fluence to the 20~keV--1~MeV BATSE fluence of $\approx5$\% (consistent with Ando et al. 2008) for the brightest available bursts (50-300~keV BATSE  fluence $>10^{-5}$~erg/cm$^2$), and much weaker, $\approx1$, for dimmer bursts. 

The roughly linear dependence of EGRET's effective area between 30~MeV and 200~MeV, and the energy independent area at higher energy, implies that EGRET's flux sensitivity is roughly energy independent between 30~MeV and 200~MeV, and falling at higher energy. Combined with the fact that the upper limits reported in Gonz\'alez S\'anchez (2005) were obtained assuming a $dN_\gamma/dE\propto E^{-2.4}$ spectrum, these upper limits apply to the 30~MeV--200~MeV fluence (the constraint implied on the fluence at higher energy is weaker and depends on the assumed spectrum). The upper limits derived in \S~\ref{sec:non_detect} using LAT non-detections on the 0.1--1~GeV fluence are similar to those obtained by EGRET on the 30~MeV--200~MeV fluence. We obtained $R^{(2)}<0.1$ for the brightest $\sim5$\% of the bursts, and $R^{(2)}<3$\% for the brightest in the sample (GRB081207 and GRB090829A), see fig.~2.

\section{Summary and discussion}
\label{sec:discussion}

We have shown that the LGRB LF inferred from the sample of bursts detected by Fermi's GBM is consistent with those determined by BATSE and {\it Swift}, see table~1, and that the GBM samples this LF in a manner similar to that of BATSE \& {\it Swift}, see figure~1. We have derived the upper limits implied by the LAT non-detections on the ratio $R$ of the 100~MeV to 1~MeV fluences, $\nu F_\nu=\nu\int dt f_\nu(t)$ for $h\nu=1,100$~MeV, during the prompt emission. The upper limits on $R$ obtained assuming $dN_\gamma/dE\propto E^{-2}$ at $E>100$~MeV (see fig.~2), also imply upper limits on the 0.1--1~GeV fluence, which is approximately given by $\ln(10)R\nu F_\nu|_{1\rm MeV}$ (the upper limit on the 1--10~GeV fluence is few times higher and depends on the assumed spectrum), and on the ratio of the 0.1--1~GeV fluence and the 0.1--1~MeV fluence, which is $\approx R$. The upper limits on $R$ are more stringent for brighter bursts (see fig.~2), with $R<\{0.1,0.3,1\}$ for $\{5,30,60\}\%$ of the bursts (see fig.~4). This implies that for most bursts the prompt $\sim1$~GeV emission may be comparable to the $\sim1$~MeV emission, but can not dominate it. For several bright bursts with reliable determination of the photon spectral index at $\sim1$~MeV, the LAT non detection implies an upper limit to the $\sim100$~MeV flux which is $<0.1$ of the flux obtained by extrapolating the $\sim1$~MeV flux to high energy (see fig.~5). Examining figs.~2 and~4, we conclude that the ratio $R$ is not universal among GRBs. The detections and non-detection upper limits imply a spread in $R$ over at least an order of magnitude. The upper limits we obtain are similar to those inferred for the fluence at lower energy, 30--200~MeV, from EGRET's non-detections of BATSE bursts (see \S~\ref{sec:EGRET}).

The upper limits on $R$ provide constraints on models for the prompt GRB emission. Models where the prompt $\sim1$~MeV emission is produced by inverse-Compton scattering of optical synchrotron photons (e.g. Stern \& Poutanen 2004, Panaitescu \& Kumar 2007), typically predict $R\ge1$. This is not supported by the data. Such models are not necessarily ruled out by the current data, as they might be modified to include a suppression of the $\sim1$~GeV flux by pair production. 
Such modification may be required for all (widely discussed) models, in which the $\sim1$~MeV power-law photon spectrum reflects the power-law energy distribution of fast cooling electrons. The suppression of the $\sim100$~MeV flux, compared to that expected from an extrapolation of the $\sim1$~MeV power-law spectrum, suggests that either the electron energy distribution does not follow a power-law over a wide energy range, or that the high energy photons are absorbed, probably by pair production. Requiring an optical depth of $\sim1$ at $100$~MeV sets an upper limit to the expansion Lorentz factor $\Gamma\lsim10^{2.5}[(L/10^{52}{\rm erg/s})/(t_v/10\,{\rm ms})]^{1/6}$ (e.g. eq. 7 of Waxman 2003). Significant compactness of the emission region has been suggested by several authors (e.g. Guetta et al. 2001, Pe'er \& Waxman 2004). The spectrum is modified in this case, compared to the optically thin case, with 100~MeV to 1~MeV flux ratios in the range $0.01\lsim R\lsim0.1$ obtained for typical parameters (e.g. figs. 8 \& 11 of Pe'er \& Waxman 2004).

\acknowledgements
This research was supported in part by ISF, AEC and Minerva grants. DG \& EP thank the Benoziyo center for Astrophysics at the Weizmann institute for hospitality during the time at which this research was initiated. We thank Nicola Omodei, David Coward for their useful advices.

\clearpage

\begin{landscape}
\begin{table}

\caption{GBM parameters for the bursts detected by Fermi.}
\begin{tabular}{llllllllllllll}
 \hline
GRB\footnotemark[1]  & $T_{90}\footnotemark[2] $ & Fluence\footnotemark[3] & PF\footnotemark[4] & Band\footnotemark[5] & Function\footnotemark[6] & $E_p$\footnotemark[7] & ~$\alpha$\footnotemark[8] & ~$\beta$\footnotemark[9] &~$\theta$\footnotemark[10] & LAT\footnotemark[11] & $\nu f_{\nu}$(1 MeV)\footnotemark[12] & $\nu f_{\nu}$(100 MeV)\footnotemark[13] &~$z$\footnotemark[14] \\ 
 & (s) & ($10^{-6}$ erg~cm$^{-2}$) & (ph~s$^{-1}$~cm$^{-2}$) & (keV) & & (keV) & & & (deg) & & (erg~s$^{-1}$~cm$^{-2}$) & (erg~s$^{-1}$~cm$^{-2}$) & \\ 
  \hline
   080810  & 122.0  & 6.90e+00  & 1.85e+00  & 50 300  & PL+HEC  & 313.5  & -0.91  & -1000  & 61  & 0  &   1.48e-08  &   0  &  3.35 \\ 
  080812  &  15.0  & -1  & -1  &          *          *  & PL+HEC  & 140.0  &  0.17  & -1000  & 71  & 0  &   0  &   0  & -1 \\ 
 080816A  &  70.0  & 1.86e+01  & 3.48e+00  & 50 300  & PL+HEC  & 146.7  & -0.57  & -1000  & 55  & 0  &   6.70e-10  &   0  & -1 \\ 
 080816B  &   5.0  & -1  & 1.38e+00  & 25 1000  & PL+HEC  & 1230.0  & -0.37  & -1000  & 70  & 0  &   0  &   0  & -1 \\ 
 080817A  &  70.0  & -1  & -1  &          *          *  &      *  &  -1  & 1000.00  & -1000  & 80  & 0  &   0  &   0  & -1 \\ 
 080817B  &   6.0  & 2.60e+00  & -1  & 25 1000  &    SPL  &  -1  & -17  & -1000  & 68  & 0  &   4.16e-07  &   0  & -1 \\ 
 080818A  &  50.0  & 2.26e+00  & -1  & 50 300  &    SPL  &  -1  & -1.57  & -1000  & 68  & 0  &   0  &   0  & -1 \\ 
 080818B  &  10.0  & 1.00e+00  & -1  & 50 300  & PL+HEC  &  80.0  & -1.30  & -1000  & 68  & 0  &   1.26e-10  &   0  & -1 \\ 
  080823  &  46.0  & 4.10e+00  & -1  & 50 300  & PL+HEC  & 164.7  & -1.20  & -1000  & 77  & 0  &   4.06e-09  &   0  & -1 \\ 
  080824  &  28.0  & 2.30e+00  & -1  & 50 300  &   Band  & 100.0  & -0.40  & -2.10  & 17  & 0  &   3.92e-08  &   2.47e-08  & -1 \\ 
 080825C  &  22.0  & 2.40e+01  & -1  & 50 300  &   Band  & 155.0  & -0.39  & -2.34  & 60  & 1  &   3.99e-07  &   8.34e-08  & -1 \\ 
  080830  &  45.0  & 4.60e+00  & -1  & 50 300  &   Band  & 154.0  & -0.59  & -1.69  & 23  & 0  &   1.13e-07  &   4.69e-07  & -1 \\ 
  080904  &  22.0  & 2.25e+00  & 3.50e+00  & 50 300  &   Band  &  35.0  &  0.00  & -2.70  & 23  & 0  &   1.29e-08  &   5.13e-10  & -1 \\ 
 080905B  & 159.0  & 4.10e-02  & 2.10e-01  & 20 1000  &    SPL  &  -1  & -1.75  & -1000  & 82  & 0  &   1.03e-10  &   0  &  2.37 \\ 
 080905C  &  28.0  & 4.60e+00  & 4.40e+00  & 25 1000  & PL+HEC  &  78.8  & -0.90  & -1000  & 108  & 0  &   3.63e-12  &   0  & -1 \\ 
 080906B  &   5.0  & 1.09e+01  & 2.20e+01  & 25 1000  &   Band  & 125.3  & -0.07  & -2.10  & 32  & 0  &   6.17e-07  &   3.89e-07  & -1 \\ 
  080912  &  17.0  & 3.30e+00  & 4.10e+00  & 25 1000  &    SPL  &  -1  & -1.74  & -1000  & 56  & 0  &   8.18e-08  &   0  & -1 \\ 
 080913B  & 140.0  & 2.20e+00  & -1  & 50 300  & PL+HEC  & 114.0  & -0.69  & -1000  & 71  & 0  &   3.19e-11  &   0  & -1 \\ 
 080916A  &  60.0  & 1.50e+01  & 4.50e+00  & 25 1000  & PL+HEC  & 109.0  & -0.90  & -1000  & 76  & 0  &   1.70e-10  &   0  &  0.69 \\ 
 080916C  & 100.9  & 2.40e+02  & 6.87e+00  & 10 10000  &   Band  & 566.0  & -0.92  & -2.28  & 48  & 1  &   5.77e-07  &   1.59e-07  &  4.2\\ 
  080920  &  85.0  & 2.40e+00  & 1.29e+00  & 25 1000  &    SPL  &  -1  & -1.42  & -1000  & 16  & 0  &   1.86e-08  &   0  & -1 \\ 
  080925  &  29.0  & 9.70e+00  & -1  & 50 300  &   Band  & 120.0  & -0.53  & -2.26  & 38  & 0  &   1.24e-07  &   3.74e-08  & -1 \\ 
  080927  &  25.0  & 5.70e+00  & 2.00e+00  & 25 1000  &    SPL  &  -1  & -1.50  & -1000  & 75  & 0  &   1.57e-07  &   0  & -1 \\ 
  080928  &  87.0  & 1.50e+00  & -1  & 50 300  &    SPL  &  -1  & -1.80  & -1000  & -1  & 0  &   0  &   0  &  1.69 \\ 
 081003C  &  67.0  & 5.40e+00  & -1  & 50 300  &    SPL  &  -1  & -1.41  & -1000  & 48  & 0  &   1.48e-07  &   0  & -1 \\ 
 081006A  &   7.0  & 7.10e-01  & -1  & 50 300  &   Band  & 1135.0  & -0.77  & -1.80  & 16  & 0  &   2.54e-07  &   6.37e-07  & -1 \\ 
 081006B  &   9.0  & 7.30e-01  & -1  & 50 300  &    SPL  &  -1  & -1.30  & -1000  & 3  & 0  &   1.85e-07  &   0  & -1 \\ 
 081007A  &  12.0  & 1.20e+00  & 2.20e+00  & 25 900  &    SPL  &  -1  & -2.10  & -1000  & 116  & 0  &   3.72e-08  &   0  &  0.53 \\ 
  081009  &  13.0  & 8.30e+00  & -1  & 8 1000  &   Band  &  20.7  &  0.20  & -4.00  & 67  & 0  &   2.53e-10  &   2.53e-14  & -1 \\ 
  081012  &  30.0  & 3.80e+00  & 2.00e+00  & 25 900  & PL+HEC  & 360.0  & -0.31  & -1000  & 61  & 0  &   5.44e-08  &   0  & -1 \\ 
  081021  &  25.0  & 5.30e+00  & 4.20e+00  & 10 1000  &   Band  & 117.0  &  0.11  & -2.80  & 125  & 0  &   2.55e-08  &   6.42e-10  & -1 \\ 
 081024C  &  65.0  & 4.00e+00  & 1.00e+00  & 50 300  &   Band  &  65.0  & -0.60  & -2.50  & 78  & 0  &   1.21e-08  &   1.21e-09  & -1 \\ 
  081025  &  45.0  & 7.10e+00  & 4.50e+00  & 8 1000  &   Band  & 200.0  &  0.15  & -2.05  & -1  & 0  &   9.87e-08  &   7.84e-08  & -1 \\ 
 081028B  &  20.0  & 2.00e+00  & 6.90e+00  & 10 1000  & PL+HEC  &  70.0  & -0.55  & -1000  & 107  & 0  &   1.63e-14  &   0  & -1 \\ 
  \hline
 \end{tabular}
 \end{table}
 \end{landscape}
 %\pagebreak
 \newpage
 \begin{landscape}
 \begin{table}
\begin{tabular}{llllllllllllll}
 \hline
GRB & $T_{90}$ & Fluence & PF & Band & Function & $E_p$ & ~$\alpha$ & ~$\beta$ & ~$\theta$ & LAT & $\nu f_{nu}$(1 MeV) & $\nu f_{nu}$(100 MeV) &~$z$ \\ 
 & (s) & ($10^{-6}$ erg~cm$^{-2}$) & (ph~s$^{-1}$~cm$^{-2}$) & (keV) & & (keV) & & & (deg) & & (erg~s$^{-1}$~cm$^{-2}$) & (erg~s$^{-1}$~cm$^{-2}$) & \\ 
  \hline 
 081101B  &   8.0  & 1.60e+01  & 1.03e+01  & 8 1000  & PL+HEC  & 550.0  & -0.62  & -1000  & 116  & 0  &   7.99e-07  &   0  & -1 \\ 
 081102A  &  88.0  & 2.10e+00  & -1  & 50 300  &   Band  &  72.0  &  0.44  & -2.36  & -1  & 0  &   1.16e-08  &   2.22e-09  &  3.04 \\ 
 081102B  &   2.2  & 1.12e+00  & 3.68e+00  & 8 1000  &    SPL  &  -1  & -17  & -1000  & 53  & 0  &   1.03e-06  &   0  & -1 \\ 
  081107  &   2.2  & 1.64e+00  & 1.10e+01  & 8 1000  &   Band  &  65.0  &  0.25  & -2.80  & 52  & 0  &   5.04e-08  &   1.27e-09  & -1 \\ 
 081109A  &  45.0  & 6.53e+00  & 3.20e+00  & 8 1000  & PL+HEC  & 240.0  & -1.28  & -1000  & -1  & 0  &   1.92e-08  &   0  & -1 \\ 
  081110  &  20.0  & -1  & -1  & 8 1000  &      *  &  -1  & 1000.00  & -1000  & 67  & 0  &   0  &   0  & -1 \\ 
 081118B  &  20.0  & 1.12e-01  & 6.70e-01  & 8 1000  &   Band  &  41.2  &  0.80  & -2.14  & 41  & 0  &   7.07e-10  &   3.71e-10  & -1 \\ 
  081120  &  12.0  & 2.70e+00  & 5.10e+00  & 8 1000  &   Band  &  44.0  &  0.40  & -2.18  & 84  & 0  &   5.05e-08  &   2.20e-08  & -1 \\ 
 081122A  &  26.0  & 9.60e+00  & 3.00e+01  & 8 1000  &   Band  & 158.6  & -0.63  & -2.24  & 21  & 0  &   8.14e-08  &   2.69e-08  & -1 \\ 
  081124  &  35.0  & 9.50e-02  & 6.70e-01  & 8 1000  &   Band  &  22.8  & -0.60  & -2.83  & 86  & 0  &   7.57e-11  &   1.66e-12  & -1 \\ 
  081125  &  15.0  & 4.91e+01  & 2.70e+01  & 8 1000  &   Band  & 221.0  &  0.14  & -2.34  & 126  & 0  &   9.23e-07  &   1.93e-07  & -1 \\ 
  081129  &  59.0  & 2.00e+01  & 1.40e+01  & 8 1000  &   Band  & 150.0  & -0.50  & -1.84  & 118  & 0  &   1.31e-07  &   2.73e-07  & -1 \\ 
 081130B  &  12.0  & 1.30e+00  & 1.80e+00  & 50 300  & PL+HEC  & 152.0  & -0.77  & -1000  & 66  & 0  &   1.10e-09  &   0  & -1 \\ 
 081204C  &   4.7  & 1.48e+00  & 7.20e+00  & 8 1000  &    SPL  &  -1  & -1.40  & -1000  & 56  & 0  &   3.03e-07  &   0  & -1 \\ 
 081206A  &  24.0  & 4.00e+00  & 2.40e+00  & 8 1000  &   Band  & 151.0  &  0.13  & -2.20  & 102  & 0  &   4.88e-08  &   1.94e-08  & -1 \\ 
 081206B  &  10.0  & -1  & -1  &          *          *  &      *  &  -1  & 1000.00  & -1000  & 82  & 0  &   0  &   0  & -1 \\ 
 081206C  &  20.0  & 1.19e+00  & 7.60e-01  & 50 300  &    SPL  &  -1  & -1.35  & -1000  & 71  & 0  &   0  &   0  & -1 \\ 
  081207  & 153.0  & 1.06e+02  & -1  & 10 1000  &   Band  & 639.0  & -0.65  & -2.41  & 56  & 0  &   8.91e-07  &   1.35e-07  & -1 \\ 
 081215A  &   7.7  & 5.44e+01  & 6.89e+01  & 8 1000  &   Band  & 304.0  & -0.58  & -2.07  & 86  & 1  &   1.31e-06  &   9.69e-07  & -1 \\ 
 081215B  &  90.0  & 2.80e+00  & -1  & 50 300  & PL+HEC  & 139.0  & -0.14  & -1000  & 112  & 0  &   8.24e-11  &   0  & -1 \\ 
  081217  &  39.0  & 1.00e+01  & 4.00e+00  & 8 1000  &   Band  & 167.0  & -0.61  & -2.70  & 54  & 0  &   7.20e-08  &   2.87e-09  & -1 \\ 
  081221  &  40.0  & 3.70e+01  & 3.30e+01  & 8 1000  &   Band  &  77.0  & -0.42  & -2.91  & 78  & 0  &   4.87e-08  &   7.37e-10  & -1 \\ 
  081222  &  30.0  & 1.35e+01  & 1.48e+01  & 8 1000  &   Band  & 134.0  & -0.55  & -2.10  & 50  & 0  &   3.08e-07  &   1.94e-07  &  2.77 \\ 
  081224  &  50.0  & -1  & -1  &          *          *  &      *  &  -1  & 1000.00  & -1000  & 16  & 0  &   0  &   0  & -1 \\ 
  081225  &  42.0  & 2.45e+00  & 6.00e-01  & 50 300  &    SPL  &  -1  & -1.51  & -1000  & 55  & 0  &   0  &   0  & -1 \\ 
 081226C  &  60.0  & 2.32e+00  & 4.50e+00  & 8 1000  & PL+HEC  &  82.0  & -14  & -1000  & 54  & 0  &   1.54e-11  &   0  & -1 \\ 
  081231  &  29.0  & 1.20e+01  & 1.53e+00  & 8 1000  &   Band  & 152.3  & -0.80  & -2.03  & 20  & 0  &   8.25e-08  &   7.19e-08  & -1 \\ 
     090107B  &  24.1  & 1.75e+00  & 3.68e+00  & 8 1000  & PL+HEC  & 106.1  & -0.68  & -1000  & -1  & 0  &   9.11e-12  &   0  & -1 \\ 
  090109  &   5.0  & 1.21e+00  & 2.76e+00  & 8 1000  &    SPL  &  -1  & -1.50  & -1000  & 62  & 0  &   1.62e-07  &   0  & -1 \\ 
 090112A  &  65.0  & 5.20e+00  & 7.00e+00  & 8 1000  &   Band  & 150.0  & -0.94  & -2.01  & 4  & 0  &   6.23e-08  &   5.95e-08  & -1 \\ 
 090112B  &  12.0  & 5.40e+00  & 1.40e+01  & 8 1000  &   Band  & 139.0  & -0.75  & -2.43  & 95  & 0  &   1.01e-07  &   1.39e-08  & -1 \\ 
 090117A  &  21.0  & 1.80e+00  & 9.60e+00  & 8 1000  &   Band  &  25.0  & -0.40  & -2.50  & 51  & 0  &   1.26e-08  &   1.26e-09  & -1 \\ 
 090117B  &  27.0  & 2.10e+00  & 4.60e+00  & 8 1000  &    SPL  &  -1  & -1.55  & -1000  & 49  & 0  &   3.95e-08  &   0  & -1 \\ 
 090117C  &  86.0  & 1.10e+01  & 4.20e+00  & 8 1000  &   Band  & 247.0  & -1  & -2.10  & 54  & 0  &   1.05e-07  &   6.65e-08  & -1 \\ 
  \hline
 \end{tabular}
 \end{table}
 \end{landscape}
 %\pagebreak
 \newpage
 \begin{landscape}
 \begin{table}
\begin{tabular}{llllllllllllll}
 \hline
GRB & $T_{90}$ & Fluence & PF & Band & Function & $E_p$ & ~$\alpha$ & ~$\beta$ & ~$\theta$ & LAT & $\nu f_{nu}$(1 MeV) & $\nu f_{nu}$(100 MeV) &~$z$ \\ 
 & (s) & ($10^{-6}$ erg~cm$^{-2}$) & (ph~s$^{-1}$~cm$^{-2}$) & (keV) & & (keV) & & & (deg) & & (erg~s$^{-1}$~cm$^{-2}$) & (erg~s$^{-1}$~cm$^{-2}$) & \\ 
  \hline 
090126B  &  10.8  & 1.25e+00  & 4.90e+00  & 8 1000  & PL+HEC  &  47.5  & -0.99  & -1000  & 18  & 0  &   1.85e-15  &   0  & -1 \\ 
  090129  &  17.2  & 5.60e+00  & 8.00e+00  & 8 1000  &   Band  & 123.2  & -1.39  & -1.98  & 22  & 0  &   9.45e-08  &   1.04e-07  & -1 \\ 
  090131  &  36.4  & 2.23e+01  & 4.79e+01  & 8 1000  &   Band  &  58.4  & -1.27  & -2.26  & 40  & 0  &   7.33e-08  &   2.21e-08  & -1 \\ 
  090202  &  66.0  & 8.65e+00  & 7.77e+00  & 8 1000  & PL+HEC  & 570.0  & -1.31  & -1000  & 55  & 0  &   1.66e-07  &   0  & -1 \\ 
  090207  &  10.0  & 4.01e+00  & 1.88e+00  & 8 1000  &    SPL  &  -1  & -1.59  & -1000  & 45  & 0  &   1.10e-07  &   0  & -1 \\ 
  090217  &  32.8  & 3.08e+01  & 1.12e+01  & 8 1000  &   Band  & 610.0  & -0.85  & -2.86  & 34  & 1  &   6.62e-07  &   9.74e-09  & -1 \\ 
  090222  &  18.0  & 2.19e+00  & 1.10e+00  & 8 1000  &   Band  & 147.9  & -0.97  & -2.56  & 80  & 0  &   1.71e-08  &   1.30e-09  & -1 \\ 
 090227A  &  50.0  & 9.00e+00  & 4.57e+00  & 8 1000  &   Band  & 1357.0  & -0.92  & -3.60  & 21  & 0  &   4.00e-07  &   4.30e-10  & -1 \\ 
 090228B  &   7.2  & 9.96e-01  & 2.53e+00  & 8 1000  & PL+HEC  & 147.8  & -0.70  & -1000  & 20  & 0  &   4.01e-10  &   0  & -1 \\ 
 090301B  &  28.0  & 2.69e+00  & 4.40e+00  & 8 1000  &   Band  & 427.0  & -0.98  & -1.93  & 56  & 0  &   2.21e-07  &   3.05e-07  & -1 \\ 
 090306C  &  38.8  & 9.00e-01  & 2.40e+00  & 8 1000  &   Band  &  87.0  & -0.32  & -2.28  & 14  & 0  &   1.52e-08  &   4.19e-09  & -1 \\ 
 090307B  &  30.0  & 1.70e+00  & 1.80e+00  & 8 1000  & PL+HEC  & 212.0  & -0.70  & -1000  & 83  & 0  &   2.71e-09  &   0  & -1 \\ 
 090308B  &   2.1  & 3.46e+00  & 1.42e+01  & 8 1000  & PL+HEC  & 710.3  & -0.54  & -1000  & 50  & 0  &   1.06e-06  &   0  & -1 \\ 
 090309B  &  60.0  & 4.70e+00  & 4.43e+00  & 8 1000  & PL+HEC  & 197.0  & -1.52  & -1000  & 26  & 0  &   1.38e-08  &   0  & -1 \\ 
  090310  & 125.2  & 2.15e+00  & 4.40e+00  & 8 1000  & PL+HEC  & 279.0  & -0.65  & -1000  & 77  & 0  &   4.02e-08  &   0  & -1 \\ 
  090319  &  67.7  & 7.47e+00  & 3.85e+00  & 8 1000  & PL+HEC  & 187.3  &  0.90  & -1000  & 27  & 0  &   4.95e-11  &   0  & -1 \\ 
 090320A  &  10.0  & -1  & -1  & 8 1000  &      *  &  -1  & 1000.00  & -1000  & 60  & 0  &   0  &   0  & -1 \\ 
 090320B  &  52.0  & 1.10e+00  & 1.20e-01  & 8 1000  & PL+HEC  &  72.0  & -1.10  & -1000  & 101  & 0  &   7.16e-13  &   0  & -1 \\ 
 090320C  &   4.0  & -1  & -1  & 8 1000  &      *  &  -1  & 1000.00  & -1000  & 40  & 0  &   0  &   0  & -1 \\ 
  090323  &  70.0  & 1.00e+02  & 1.23e+01  & 8 1000  & PL+HEC  & 697.0  & -0.89  & -1000  & -1  & 1  &   6.46e-07  &   0  &  3.57 \\ 
  090326  &  11.2  & 8.60e-01  & -1  & 8 1000  & PL+HEC  &  75.0  & -0.86  & -1000  & 103  & 0  &   4.17e-13  &   0  & -1 \\ 
  090327  &  24.0  & 3.00e+00  & 3.50e+00  & 8 1000  &   Band  &  89.7  & -0.39  & -2.90  & 66  & 0  &   1.09e-08  &   1.72e-10  & -1 \\ 
 090328A  & 100.0  & 8.09e+01  & 1.85e+01  & 8 1000  &   Band  & 653.0  & -0.93  & -2.20  & -1  & 1  &   1.40e-06  &   5.57e-07  &  0.74 \\ 
  090330  &  80.0  & 1.14e+01  & 6.80e+00  & 8 1000  &   Band  & 246.0  & -0.99  & -2.68  & 50  & 0  &   7.45e-08  &   3.25e-09  & -1 \\ 
  090403  &  16.0  & -1  & -1  &          *          *  &      *  &  -1  & 1000.00  & -1000  & 42  & 0  &   0  &   0  & -1 \\ 
  090409  &  20.0  & 6.14e-01  & 1.36e+00  & 8 1000  & PL+HEC  & 137.0  &  1.20  & -1000  & 90  & 0  &   8.53e-14  &   0  & -1 \\ 
 090411A  &  24.6  & 8.60e+00  & 3.25e+00  & 8 1000  &   Band  & 141.0  & -0.88  & -1.82  & 59  & 0  &   1.57e-07  &   3.60e-07  & -1 \\ 
 090411B  &  18.7  & 8.00e+00  & 7.40e+00  & 8 1000  &   Band  & 189.0  & -0.80  & -2.00  & 111  & 0  &   1.09e-07  &   1.09e-07  & -1 \\ 
  090422  &  10.0  & 1.00e+00  & 7.80e+00  & 8 1000  &    SPL  &  -1  &  1.81  & -1000  & 29  & 0  &   2.89e-07  &   0  & -1 \\ 
  090423  &  12.0  & 1.10e+00  & 3.30e+00  & 8 1000  & PL+HEC  &  82.0  & -0.77  & -1000  & 75.6  & 0  &   8.96e-13  &   0  &  8.10 \\ 
  090424  &  52.0  & 5.20e+01  & 1.37e+02  & 8 1000  &   Band  & 177.0  &  0.90  & -2.90  & 71  & 0  &   1.13e-07  &   1.79e-09  &  0.54 \\ 
  090425  &  72.0  & 1.30e+01  & 1.40e+01  & 8 1000  &   Band  &  69.0  & -1.29  & -2.03  & 105  & 0  &   1.28e-07  &   1.12e-07  & -1 \\ 
 090426B  &   3.8  & 5.20e-01  & -1  & 8 1000  &    SPL  &  -1  & -1.60  & -1000  & 56  & 0  &   4.75e-08  &   0  & -1 \\ 
 090426C  &  12.0  & 3.10e+00  & 6.80e+00  & 8 1000  &   Band  & 295.0  & -1.29  & -1.98  & 69  & 0  &   9.65e-08  &   1.06e-07  & -1 \\ 
  \hline
 \end{tabular}
 \end{table}
 \end{landscape}
 %\pagebreak
 \newpage
 \begin{landscape}
 \begin{table}
\begin{tabular}{llllllllllllll}
 \hline
GRB & $T_{90}$ & Fluence & PF & Band & Function & $E_p$ & ~$\alpha$ & ~$\beta$ & ~$\theta$ & LAT & $\nu f_{nu}$(1 MeV) & $\nu f_{nu}$(100 MeV) &~$z$ \\ 
 & (s) & ($10^{-6}$ erg~cm$^{-2}$) & (ph~s$^{-1}$~cm$^{-2}$) & (keV) & & (keV) & & & (deg) & & (erg~s$^{-1}$~cm$^{-2}$) & (erg~s$^{-1}$~cm$^{-2}$) & \\ 
  \hline 
 090427B  &   7.0  & 8.00e-01  & -1  & 8 1000  &    SPL  &  -1  & -1.10  & -1000  & 14  & 0  &   1.19e-07  &   0  & -1 \\ 
 090427C  &  12.5  & 1.60e+00  & -1  & 8 1000  & PL+HEC  &  75.0  &  0.35  & -1000  & 81  & 0  &   7.59e-18  &   0  & -1 \\ 
 090428A  &   8.0  & 9.90e-01  & 1.23e+01  & 8 1000  &   Band  &  85.0  & -0.40  & -2.70  & 96  & 0  &   3.13e-08  &   1.25e-09  & -1 \\ 
 090428B  &  30.0  & 5.20e+00  & 1.01e+01  & 8 1000  &   Band  &  53.0  & -1.81  & -2.17  & 101  & 0  &   3.25e-08  &   1.49e-08  & -1 \\ 
 090429C  &  13.0  & 3.70e+00  & 6.70e+00  & 8 1000  &    SPL  &  -1  & -1.43  & -1000  & 112  & 0  &   1.73e-07  &   0  & -1 \\ 
 090429D  &  11.0  & 1.60e+00  & 8.60e-01  & 8 1000  &    SPL  & 223.0  & -0.87  & -1000  & 33  & 0  &   1.27e-07  &   0  & -1 \\ 
  090502  &  66.2  & 3.50e-02  & 6.20e+00  & 8 1000  & PL+HEC  &  63.2  & -1.10  & -1000  & 77  & 0  &   3.68e-15  &   0  & -1 \\ 
  090509  & 295.0  & 8.40e+00  & 3.10e+00  & 8 1000  &    SPL  &  -1  & -1.55  & -1000  & 75  & 0  &   2.18e-10  &   0  & -1 \\ 
 090510B  &   7.0  & -1  & -1  &          *          *  &      *  &  -1  & 1000.00  & -1000  & 100  & 0  &   0  &   0  & -1 \\ 
  090511  &  14.0  & 1.80e+00  & 2.50e+00  & 8 1000  & PL+HEC  & 391.0  & -0.95  & -1000  & 67  & 0  &   4.26e-08  &   0  & -1 \\ 
 090513A  &  23.0  & 6.80e+00  & 2.70e+00  & 8 1000  & PL+HEC  & 850.0  & -0.90  & -1000  & 89  & 0  &   1.91e-07  &   0  & -1 \\ 
  090514  &  49.0  & 8.10e+00  & 7.60e+00  & 8 1000  &    SPL  &  -1  & -1.92  & -1000  & 19  & 0  &   3.47e-08  &   0  & -1 \\ 
 090516A  & 350.0  & 2.30e+01  & 5.30e+00  & 8 1000  &   Band  &  51.4  & -13  & -2.10  & 20  & 0  &   1.49e-08  &   9.41e-09  &  4.11 \\ 
 090516B  & 350.0  & 3.00e+01  & 4.00e+00  & 8 1000  & PL+HEC  & 327.0  & -11  & -1000  & 45  & 0  &   3.38e-08  &   0  & -1 \\ 
 090516C  &  15.0  & 4.00e+00  & 7.70e+00  & 8 1000  &   Band  &  38.0  & -0.44  & -1.81  & 69  & 0  &   8.74e-08  &   2.10e-07  & -1 \\ 
 090518A  &   9.0  & 1.60e+00  & 4.70e+00  & 8 1000  &    SPL  &  -1  & -1.59  & -1000  & 53  & 0  &   8.38e-08  &   0  & -1 \\ 
 090518B  &  12.0  & 2.20e+00  & 5.60e+00  & 8 1000  & PL+HEC  & 127.0  & -0.74  & -1000  & 90  & 0  &   3.00e-10  &   0  & -1 \\ 
 090519B  &  87.0  & 1.40e+00  & 5.02e+00  & 8 1000  &    SPL  &  -1  & -1.63  & -1000  & 18  & 0  &   1.74e-07  &   0  & -1 \\ 
 090520C  &   4.9  & 3.54e+00  & 4.47e+00  & 8 1000  &   Band  & 204.2  & -0.73  & -1.96  & 71  & 0  &   2.75e-07  &   3.31e-07  & -1 \\ 
 090520D  &  12.0  & 4.00e+00  & 4.10e+00  & 8 1000  &   Band  &  46.3  & -0.99  & -3.25  & 66  & 0  &   3.32e-09  &   1.05e-11  & -1 \\ 
  090522  &  22.0  & 1.20e+00  & 3.50e+00  & 8 1000  & PL+HEC  &  75.8  & -13  & -1000  & 53  & 0  &   1.97e-12  &   0  & -1 \\ 
  090524  &  72.0  & 1.85e+01  & 1.41e+01  & 8 1000  &   Band  &  82.6  & -1  & -2.30  & 63  & 0  &   3.86e-08  &   9.70e-09  & -1 \\ 
 090528A  &  68.0  & 9.30e+00  & 7.60e+00  & 8 1000  & PL+HEC  &  99.0  & -1.70  & -1000  & 81  & 0  &   9.53e-09  &   0  & -1 \\ 
 090528B  & 102.0  & 4.65e+01  & 1.47e+01  & 8 1000  &   Band  & 172.0  & -1.10  & -2.30  & 65  & 0  &   1.56e-07  &   3.91e-08  & -1 \\ 
 090529B  &   5.1  & 3.40e-01  & 4.10e+00  & 8 1000  &   Band  & 142.0  & -0.70  & -2.00  & 36  & 0  &   1.42e-08  &   1.42e-08  & -1 \\ 
 090529C  &  10.4  & 3.10e+00  & 2.50e+01  & 8 1000  &   Band  & 188.0  & -0.84  & -2.10  & 69  & 0  &   8.64e-08  &   5.45e-08  & -1 \\ 
 090530B  & 194.0  & 5.90e+01  & 1.08e+01  & 8 1000  &   Band  &  67.0  & -0.71  & -2.42  & 84  & 0  &   5.92e-08  &   8.55e-09  & -1 \\ 
090602  &  16.0  & 5.70e+00  & 3.62e+00  & 8 1000  & PL+HEC  & 503.0  & -0.56  & -1000  & 112  & 0  &   1.84e-07  &   0  & -1 \\ 
  090606  &  60.0  & 3.19e+00  & 2.41e+00  & 8 1000  &    SPL  &  -1  & -1.63  & -1000  & 128  & 0  &   8.15e-08  &   0  & -1 \\ 
  090608  &  61.0  & 3.20e+00  & 2.70e+00  & 8 1000  &    SPL  &  -1  & -1.83  & -1000  & 93  & 0  &   2.50e-08  &   0  & -1 \\ 
 090610A  &   6.5  & 7.32e-01  & 9.40e-01  & 8 1000  &    SPL  &  -1  & -1.30  & -1000  & 70  & 0  &   8.42e-08  &   0  & -1 \\ 
 090610B  & 202.5  & 4.13e+00  & 1.54e+00  & 8 1000  &    SPL  &  -1  & -1.66  & -1000  & 91  & 0  &   8.41e-10  &   0  & -1 \\ 
 090610C  &  18.1  & 8.54e-01  & 1.12e+00  & 8 1000  &    SPL  &  -1  & -1.62  & -1000  & 104  & 0  &   3.14e-08  &   0  & -1 \\ 
  090612  &  58.0  & 2.37e+00  & 1.63e+00  & 8 1000  &   Band  & 357.0  & -0.60  & -1.90  & 56  & 0  &   1.75e-07  &   2.78e-07  & -1 \\ 
  \hline
 \end{tabular}
 \end{table}
 \end{landscape}
% \pagebreak
 \newpage
 \begin{landscape}
 \begin{table}
\begin{tabular}{llllllllllllll}
 \hline
GRB & $T_{90}$ & Fluence & PF & Band & Function & $E_p$ & $~\alpha$ & ~$\beta$ & ~$\theta$ & LAT & $\nu f_{nu}$(1 MeV) & $\nu f_{nu}$(100 MeV) & ~$z$ \\ 
 & (s) & ($10^{-6}$ erg~cm$^{-2}$) & (ph~s$^{-1}$~cm$^{-2}$) & (keV) & & (keV) & & & (deg) & & (erg~s$^{-1}$~cm$^{-2}$) & (erg~s$^{-1}$~cm$^{-2}$) & \\ 
  \hline 
  090616  &   2.7  & 2.23e-01  & 2.08e+00  & 8 1000  &    SPL  &  -1  & -1.27  & -1000  & 68  & 0  &   2.62e-07  &   0  & -1 \\ 
  090618  & 155.0  & 2.70e+02  & 7.34e+01  & 8 1000  &   Band  & 155.5  & -1.26  & -2.50  & 133  & 0  &   2.75e-07  &   2.75e-08  &  0.54 \\ 
  090620  &  16.5  & 6.60e+00  & 7.00e+00  & 8 1000  &   Band  & 156.0  & -0.40  & -2.44  & 60  & 0  &   7.19e-08  &   9.48e-09  & -1 \\ 
 090621A  & 294.0  & 4.40e+00  & 1.92e+00  & 8 1000  &   Band  &  56.0  & -1.10  & -2.12  & 12  & 0  &   1.59e-08  &   9.15e-09  & -1 \\ 
 090621C  &  59.9  & 1.80e+00  & 2.29e+00  & 8 1000  & PL+HEC  & 148.0  & -1.40  & -1000  & 52  & 0  &   1.79e-09  &   0  & -1 \\ 
 090621D  &  39.9  & 1.34e+00  & 1.74e+00  & 8 1000  &    SPL  &  -1  & -1.66  & -1000  & 79  & 0  &   2.63e-08  &   0  & -1 \\ 
  090623  &  72.2  & 9.60e+00  & 3.30e+00  & 8 1000  &   Band  & 428.0  & -0.69  & -2.30  & 73  & 0  &   8.35e-08  &   2.10e-08  & -1 \\ 
 090625A  &  51.0  & 8.80e-01  & 5.00e-01  & 8 1000  & PL+HEC  & 198.0  & -0.60  & -1000  & 13  & 0  &   7.18e-10  &   0  & -1 \\ 
 090625B  &  13.6  & 1.04e+00  & 1.87e+00  & 8 1000  &   Band  & 100.0  & -0.40  & -2.00  & 125  & 0  &   3.62e-08  &   3.62e-08  & -1 \\ 
  090626  &  70.0  & 3.50e+01  & 1.79e+01  & 8 1000  &   Band  & 175.0  & -1.29  & -1.98  & 15  & 1  &   1.43e-07  &   1.57e-07  & -1 \\ 
  090630  &   5.1  & 5.10e-01  & 2.78e+00  & 8 1000  &   Band  &  71.0  & -1.50  & -2.30  & 75  & 0  &   1.32e-08  &   3.30e-09  & -1 \\ 
  090701  &  12.0  & 4.50e-01  & 2.10e+00  & 8 1000  &    SPL  &  -1  &  1.84  & -1000  & 13  & 0  &   1.44e-07  &   0  & -1 \\ 
  090703  &   9.0  & 6.80e-01  & 1.00e+00  & 8 1000  &    SPL  &  -1  & -1.72  & -1000  & 25  & 0  &   2.76e-08  &   0  & -1 \\ 
  090704  &  70.0  & 5.80e+00  & 1.20e+00  & 8 1000  & PL+HEC  & 233.7  & -1.13  & -1000  & 77  & 0  &   6.15e-09  &   0  & -1 \\ 
  090706  & 100.0  & 1.50e+00  & 1.24e+00  & 8 1000  &    SPL  &  -1  & -2.16  & -1000  & 20  & 0  &   4.90e-09  &   0  & -1 \\ 
  090708  &  18.0  & 4.00e-01  & 1.00e+00  & 8 1000  & PL+HEC  &  47.5  & -1.29  & -1000  & 55  & 0  &   4.90e-14  &   0  & -1 \\ 
 090709B  &  32.0  & 1.30e+00  & 2.00e+00  & 8 1000  & PL+HEC  & 130.0  & -11  & -1000  & 35  & 0  &   1.71e-10  &   0  & -1 \\ 
  090711  & 100.0  & 1.17e+01  & 4.20e+00  & 8 1000  & PL+HEC  & 210.0  & -1.30  & -1000  & 13  & 0  &   8.06e-09  &   0  & -1 \\ 
  090712  &  72.0  & 4.20e+00  & 6.30e-01  & 8 1000  & PL+HEC  & 505.0  & -0.68  & -1000  & 33  & 0  &   1.96e-08  &   0  & -1 \\ 
  090713  & 113.0  & 3.70e+00  & 1.60e+00  & 8 1000  & PL+HEC  &  99.0  & -0.34  & -1000  & 63  & 0  &   3.98e-13  &   0  & -1 \\ 
 090717A  &  70.0  & 4.50e-01  & 7.80e+00  & 8 1000  &   Band  & 120.0  & -0.88  & -2.33  & 70  & 0  &   4.66e-09  &   1.02e-09  & -1 \\ 
 090718B  &  28.0  & 2.52e+01  & 3.20e+01  & 8 1000  &   Band  & 184.0  & -1.18  & -2.59  & 76  & 0  &   1.24e-07  &   8.19e-09  & -1 \\ 
  090719  &  16.0  & 4.83e+01  & 3.78e+01  & 8 1000  &   Band  & 254.0  & -0.68  & -2.92  & 88  & 0  &   4.55e-07  &   6.57e-09  & -1 \\ 
 090720A  &   7.0  & 2.90e+00  & 1.09e+01  & 8 1000  & PL+HEC  & 117.5  & -0.75  & -1000  & 113  & 0  &   2.14e-10  &   0  & -1 \\ 
 090720B  &  20.0  & 1.06e+01  & 1.09e+01  & 8 1000  &   Band  & 924.0  & -1  & -2.43  & 56  & 0  &   2.96e-07  &   3.97e-08  & -1 \\ 
 090807B  &   3.0  & 1.02e+00  & 1.09e+01  & 8 1000  &   Band  &  37.0  & -0.60  & -2.40  & 45  & 0  &   4.65e-08  &   7.38e-09  & -1 \\ 
 090809B  &  15.0  & 2.26e+01  & 2.36e+01  & 8 1000  &   Band  & 198.0  & -0.85  & -2.02  & 81  & 0  &   4.95e-07  &   4.51e-07  & -1 \\ 
  090813  &   9.0  & 3.50e+00  & 1.44e+01  & 8 1000  &   Band  &  95.0  & -1.25  & -2.00  & 35.3  & 0  &   9.45e-08  &   9.45e-08  & -1 \\ 
 090815A  & 200.0  & 3.40e+00  & 1.90e+00  & 8 1000  &    SPL  &  -1  & -1.50  & -1000  & 87  & 0  &   6.29e-08  &   0  & -1 \\ 
 090815B  &  30.0  & 5.05e+00  & 1.44e+01  & 8 1000  &   Band  &  18.6  & -1.82  & -2.70  & 82  & 0  &   6.71e-09  &   2.67e-10  & -1 \\ 
  090817  & 220.0  & 7.30e+00  & 3.80e+00  & 8 1000  &   Band  & 115.0  & -1.10  & -2.20  & 82  & 0  &   6.47e-09  &   2.58e-09  & -1 \\ 
 090820A  &  60.0  & 6.60e+01  & 5.80e+01  & 8 1000  &   Band  & 215.0  & -0.69  & -2.61  & 108  & 0  &   3.64e-07  &   2.19e-08  & -1 \\ 
 \hline
 \end{tabular}
 \end{table}
 \end{landscape}
% \pagebreak
 \newpage
 \begin{landscape}
 \begin{table}
\begin{tabular}{llllllllllllll}
 \hline
GRB & $T_{90}$ & Fluence & PF & Band & Function & $E_p$ & $~\alpha$ & ~$\beta$ & ~$\theta$ & LAT & $\nu f_{nu}$(1 MeV) & $\nu f_{nu}$(100 MeV) & ~$z$ \\ 
 & (s) & ($10^{-6}$ erg~cm$^{-2}$) & (ph~s$^{-1}$~cm$^{-2}$) & (keV) & & (keV) & & & (deg) & & (erg~s$^{-1}$~cm$^{-2}$) & (erg~s$^{-1}$~cm$^{-2}$) & \\ 
  \hline 
 090820B  &  11.2  & 1.16e+00  & 6.10e+00  & 8 1000  & PL+HEC  &  38.8  & -1.44  & -1000  & 32  & 0  &   2.10e-13  &   0  & -1 \\ 
  090826  &   8.5  & 1.26e+00  & 3.28e+00  & 8 1000  & PL+HEC  & 172.0  & -0.96  & -1000  & 35  & 0  &   1.55e-09  &   0  & -1 \\ 
  090828  & 100.0  & 2.52e+01  & 1.62e+01  & 8 1000  &   Band  & 136.5  & -1.23  & -2.12  & 95  & 0  &   5.50e-08  &   3.16e-08  & -1 \\ 
 090829A  &  85.0  & 1.02e+02  & 5.15e+01  & 8 1000  &   Band  & 183.0  & -1.44  & -2.10  & 47  & 0  &   2.44e-07  &   1.54e-07  & -1 \\ 
 090829B  & 100.0  & 6.40e+00  & 3.20e+00  & 8 1000  &   Band  & 143.0  & -0.70  & -2.40  & 42  & 0  &   3.39e-08  &   5.38e-09  & -1 \\ 
  090831  &  69.1  & 1.66e+01  & 9.40e+00  & 8 1000  &   Band  & 243.8  & -1.52  & -1.96  & 107  & 0  &   9.92e-08  &   1.19e-07  & -1 \\ 
 090902B  &  21.0  & 3.74e+02  & 4.61e+01  & 50 10000  &   Band  & 798.0  & -0.61  & -3.87  & 52  & 1  &   7.16e-06  &   8.88e-10  &  1.82 \\ 
 090904B  &  71.0  & 2.44e+01  & 9.80e+00  & 8 1000  &   Band  & 106.3  & -1.26  & -2.18  & 113  & 0  &   6.20e-08  &   2.71e-08  & -1 \\ 
  090910  &  62.0  & 9.20e+00  & 2.30e+00  & 8 1000  &   Band  & 274.8  & -0.90  & -2.00  & 107  & 0  &   6.32e-08  &   6.32e-08  & -1 \\ 
 090922A  &  92.0  & 1.14e+01  & 1.56e+01  & 8 1000  &   Band  & 139.3  & -0.77  & -2.28  & 19  & 0  &   2.22e-07  &   6.11e-08  & -1 \\ 
  090925  &  50.0  & 9.46e+00  & 4.20e+00  & 8 1000  &   Band  & 156.0  & -0.60  & -1.91  & 116  & 0  &   1.59e-07  &   2.41e-07  & -1 \\ 
 090926A  &  20.0  & 1.45e+02  & 8.08e+01  & 8 1000  &   Band  & 268.0  & -0.69  & -2.34  & 52  & 1  &   1.85e-06  &   3.84e-07  &  2.11 \\ 
 090926B  &  81.0  & 8.70e+00  & -1  & 10 1000  & PL+HEC  &  91.0  & -0.13  & -1000  & 100  & 0  &   6.54e-14  &   0  &  1.24 \\ 
 090929A  &   8.5  & 1.06e+01  & 1.09e+01  & 8 1000  & PL+HEC  & 610.9  & -0.52  & -1000  & 122  & 0  &   9.81e-07  &   0  & -1 \\ 
 091003A  &  21.1  & 3.76e+01  & 3.18e+01  & 8 1000  &   Band  & 486.2  & -1.13  & -2.64  & 13  & 1  &   4.93e-07  &   2.59e-08  &  0.90 \\ 
  091010  &   8.1  & 1.09e+01  & 4.09e+01  & 8 1000  & PL+HEC  & 150.0  & -1.11  & -1000  & 55.7  & 0  &   1.72e-08  &   0  & -1 \\ 
  091020  &  37.0  & 1.00e+01  & 7.40e+00  & 8 1000  &   Band  &  47.9  &  0.20  & -1.70  & 118  & 0  &   1.47e-07  &   5.84e-07  &  1.71 \\ 
  091024  & 1080.0  & -1  & -1  &          *          *  &      *  & 400.0  & 1000.00  & -1000  & 14  & 0  &   0  &   0  &  1.09 \\ 
  091030  & 160.0  & 3.03e+01  & 9.58e+00  & 8 1000  &   Band  & 507.0  & -0.88  & -2.20  & 100  & 0  &   2.92e-07  &   1.16e-07  & -1 \\ 
  091031  &  35.0  & 2.05e+01  & 7.50e+00  & 8 1000  &   Band  & 503.1  & -0.91  & -2.34  & 22  & 1  &   2.09e-07  &   4.36e-08  & -1 \\ 
 091102A  &   7.3  & 2.10e+00  & 2.90e+00  & 8 1000  &    SPL  &  -1  & -1.24  & -1000  & 94  & 0  &   3.20e-07  &   0  & -1 \\ 
  091112  &  40.0  & 9.70e+00  & -1  & 10 1000  & PL+HEC  & 750.0  & -1.13  & -1000  & 82  & 0  &   1.80e-07  &   0  & -1 \\ 
  091120  &  52.0  & 3.02e+01  & 2.13e+01  & 8 1000  &   Band  & 124.0  & -1.15  & -2.98  & 45  & 0  &   3.50e-08  &   3.83e-10  & -1 \\ 
  091123  & 650.0  & 4.07e+01  & 6.10e+00  & 8 1000  & PL+HEC  & 101.3  & -18  & -1000  & 106  & 0  &   5.65e-11  &   0  & -1 \\ 
  091127  &   9.0  & 1.87e+01  & 4.69e+01  & 8 1000  &   Band  &  36.0  & -1.27  & -2.20  & 25  & 0  &   2.64e-07  &   1.05e-07  &  0.49 \\ 
  091128  &  97.0  & 3.76e+01  & 9.30e+00  & 8 1000  &   Band  & 177.4  & -0.99  & -3.90  & 96  & 0  &   1.83e-08  &   2.89e-12  & -1 \\ 
 091208B  &  15.0  & 5.80e+00  & 3.24e+01  & 8 1000  &   Band  & 124.0  & -1.44  & -2.32  & 56  & 0  &   6.13e-08  &   1.40e-08  &  1.06 \\ 
  091221  &  32.0  & 1.38e+01  & 5.10e+00  & 8 1000  &   Band  & 207.0  & -0.69  & -2.30  & 53  & 0  &   1.03e-07  &   2.59e-08  & -1 \\ 
  090530B  & 194.0  & 5.90e+01  & 1.08e+01  & 8 1000  &   Band  &  67.0  & -0.71  & -2.42  & 84  & 0  &   5.92e-08  &   8.55e-09  & -1 \\ 
\hline
 \end{tabular}
 \end{table}
 \end{landscape}
 %\pagebreak
 \newpage
 \begin{landscape}
 \begin{table}
\begin{tabular}{llllllllllllll}
 \hline
GRB & $T_{90}$ & Fluence & PF & Band & Function & $E_p$ & $~\alpha$ & ~$\beta$ & ~$\theta$ & LAT & $\nu f_{nu}$(1 MeV) & $\nu f_{nu}$(100 MeV) & ~$z$ \\ 
 & (s) & ($10^{-6}$ erg~cm$^{-2}$) & (ph~s$^{-1}$~cm$^{-2}$) & (keV) & & (keV) & & & (deg) & & (erg~s$^{-1}$~cm$^{-2}$) & (erg~s$^{-1}$~cm$^{-2}$) & \\ 
  \hline 
100111A  &  12.0  & 1.50e+00  & 3.50e+00  & 8 1000  &    SPL  &  -1  & -1.66  & -1000  & 32  & 0  &   7.71e-08  &   0  & -1 \\ 
 100116A  & 110.0  & 3.36e+01  & -1  & 8 10000  & PL+HEC  & 1240.0  & -12  & -1000  & 29  & 1  &   6.39e-10  &   0  & -1 \\ 
 100122A  &   6.6  & 1.00e+01  & 1.04e+01  & 8 1000  &   Band  &  45.6  & -0.98  & -2.31  & 45  & 0  &   4.18e-08  &   1.00e-08  & -1 \\ 
 100130A  & 106.0  & 8.21e+00  & 5.90e+00  & 8 1000  & PL+HEC  & 100.5  & -0.97  & -1000  & 51  & 0  &   3.36e-11  &   0  & -1 \\ 
 100130B  &  90.0  & 1.34e+01  & 3.72e+00  & 8 1000  & PL+HEC  & 208.0  & -1.22  & -1000  & 89  & 0  &   7.89e-09  &   0  & -1 \\ 
 100131A  &   6.2  & 7.72e+00  & 3.38e+01  & 8 1000  &   Band  & 132.1  & -0.63  & -2.21  & 27  & 0  &   3.79e-07  &   1.44e-07  & -1 \\ 
 100205B  &  13.6  & 1.41e+00  & 2.98e+00  & 8 1000  & PL+HEC  & 124.2  & -0.47  & -1000  & 102  & 0  &   4.12e-11  &   0  & -1 \\ 
 100212A  &   2.3  & 3.81e-01  & 3.16e+00  & 8 1000  & PL+HEC  & 159.3  & -1.15  & -1000  & 15  & 0  &   9.60e-09  &   0  & -1 \\ 
 100218A  &  30.8  & 2.58e+00  & 1.40e+00  & 8 1000  &   Band  & 131.6  & -0.14  & -2.00  & 37  & 0  &   2.50e-08  &   2.50e-08  & -1 \\ 
\hline
 \end{tabular}
 \tablefoot{
\tablefoottext{1}{GRB name.}\\
\tablefoottext{2}{The duration of the GRB.}\\
\tablefoottext{3}{GRB fluence in the energy interval specified in Col. 5 (set to -1 when value is not available)}\\
\tablefoottext{4}{GRB peak flux (set to -1 when value is not available)}\\
 \tablefoottext{5}{Energy band for the fluence determination}\\
\tablefoottext{6}{Method used to fit the spectra (Band=broken power law, SPL=single power law, PL+HEC=power law and exponential cutoff)}\\
 \tablefoottext{7}{Peak energy (set to -1 when not available)}\\
 \tablefoottext{8}{The $\alpha$ spectral index (set to 1000 when not available). }\\
 \tablefoottext{9}{The $\beta$ spectral index (set to -1000 when not available). }\\
\tablefoottext{10}{The angle, $\theta$,  from the LAT boresight, in deg (set to -1 when not available)}\\
\tablefoottext{11}{LAT detection (1 = YES, 0 = NO)}\\
 \tablefoottext{12}{ The flux at $\sim$ MeV (set zero when not available). }\\
 \tablefoottext{13}{ The flux at $\sim$ 100 MeV  obtained by extrapolating to high energy the $\sim1$~MeV spectrum (set zero when not available).}\\
  \tablefoottext{14}{The redshift (equal to -1 if not measured)}
}
 
 \end{table}
\end{landscape}
\end{document}